# Vaccine Search Patterns Provide Insights into Vaccination Intent

Malahy S,*[1] Sun M,[1] Spangler KR,[2] Leibler JH,[2] Lane KJ,[2] Bavadekar S,[1] Kamath C,[1] Kumok A,[1] Sun Y,[2] Gupta J,[1] Griffith T,[1] Boulanger A,[1] Young M,[1] Stanton C,[1] Mayer Y,[1] Smith K,[1] Shekel T,[1] Chou K,[1] Corrado G,[1] Levy JI,[2] Szpiro AA,[3] Gabrilovich E,[1] Wellenius GA.[2]

\* = corresponding author, seanmalahy@google.com

[1]Google, LLC, Mountain View, CA
[2]Department of Environmental Health, Boston University School of Public Health, Boston, MA
[3]Department of Biostatistics, University of Washington School of Public Health, Seattle, WA

These authors contributed equally:
- S Malahy and M Sun
- E Gabrilovich and GA Wellenius

# Abstract

Despite ample supply of COVID-19 vaccines, the proportion of fully vaccinated individuals remains suboptimal across much of the US. Rapid vaccination of additional people will prevent new infections among both the unvaccinated and the vaccinated, thus saving lives. With the rapid rollout of vaccination efforts this year, the internet has become a dominant source of information about COVID-19 vaccines, their safety and efficacy, and their availability. We sought to evaluate whether trends in internet searches related to COVID-19 vaccination — as reflected by Google's Vaccine Search Insights (VSI) index — could be used as a marker of population-level interest in receiving a vaccination. We found that between January and August of 2021: 1) Google's weekly VSI index was associated with the number of new vaccinations administered in the subsequent three weeks, and 2) the average VSI index in earlier months was strongly correlated (up to r=0.89) with vaccination rates many months later. Given these results, we illustrate an approach by which data on search interest may be combined with other available data to inform local public health outreach and vaccination efforts. These results suggest that the VSI index may be useful as a leading indicator of population-level interest in or intent to obtain a COVID-19 vaccine, especially early in the vaccine deployment efforts. These results may be relevant to current efforts to administer COVID-19 vaccines to unvaccinated individuals, to newly eligible children, and to those eligible to receive a booster shot. More broadly, these results highlight the opportunities for anonymized and aggregated internet search data, available in near real-time, to inform the response to public health emergencies.

After reaching a minimum in June 2021, coronavirus disease 2019 (COVID-19) cases and hospitalizations rose rapidly throughout the summer across the United States as the more contagious Delta variant became the dominant strain of SARS-CoV-2, the virus that causes COVID-19. The vast majority of new hospitalizations across the US in mid-to-late 2021 have been occurring among unvaccinated individuals.[1,2] Despite ample supply of COVID-19 vaccines, the proportion of fully vaccinated individuals remains below recommended levels across much of the US. Specifically, as of October 20, 2021, the US Centers for Disease Control and Prevention (CDC) reports that only 57% of Americans (67% of the population ≥ 12 years of age) have been fully vaccinated (https://covid.cdc.gov/covid-data-tracker). Rapid vaccination of additional people will prevent new infections among both the unvaccinated and the vaccinated, reduce the severity of infections among the vaccinated, and thus, save lives.

With the rapid roll out of vaccination efforts this year, the internet has become a dominant source of information (and misinformation[3]) about COVID-19 vaccines, their safety and efficacy, and their availability. Prior studies have shown that internet search patterns based on anonymized and aggregated data can be used to predict the occurrence of Lyme disease and outbreaks of influenza; to nowcast COVID-19 cases, hospitalizations, and deaths; and to identify food establishments that would benefit from food safety inspections to limit the further spread of foodborne illness.[4–7] Internet search patterns may similarly provide novel insights that could be used to inform public health efforts to increase vaccination uptake, but this hypothesis has not been examined in detail. Internet searches related to COVID-19 vaccines began rising in January 2021 and then rose further starting in March.[8] Early evidence suggests that internet search activity (aggregated to the state level) is associated with higher rates of vaccination in that state.[9]

Google recently began publishing the COVID-19 Vaccination Search Insights (VSI) index, a publicly available dataset showing trends in Google searches related to COVID-19 vaccination from January 2021 through the present. We sought to evaluate whether patterns in internet searches across locations and over time could be used as a marker of population-level interest in receiving a vaccination and, if so, to explore how this information might be used by public health officials to identify geographic areas with particularly high amenability towards vaccination despite low uptake.

## Methods

*Search data*

We obtained publicly available data on the relative volume of searches related to COVID-19 vaccinations from Google's COVID-19 Vaccination Search Insights (VSI) (https://google-research.github.io/vaccination-search-insights). The main VSI index reflects the weekly proportion of searches in a given geographic area related to COVID-19 vaccination, indicating overall search interest in the topic. The VSI index is calculated as the weekly number of relevant searches within a given geographic region, normalized by the search activity in that region. The anonymization process for the VSI data is based on differential privacy and documented in detail elsewhere.[10] Importantly, the anonymization process limits the contribution to the VSI index of any one user such that searches for information by scientists, medical professionals, and other experts will have limited impact on the overall VSI index. The normalization procedure allows comparisons across locations and within the same location across time. We obtained and examined VSI data at the state, county, and ZIP code levels for the weeks beginning January 11, 2021 through September 6, 2021.

*Vaccination data*

We used multiple data sources to estimate the number of vaccinations administered within each US state, county, or ZIP code, since no single data source provides vaccination data at all three spatial scales for the entire time period of interest. We obtained state-level data from Our World in Data's (OWD)[11] daily measure of the cumulative number of individuals in each state with 1 or more vaccine doses for the weeks beginning January 11, 2021 through September 6, 2021, inclusive. We imputed the value for days with missing data by assuming a linear trend from the closest previous date without missing data to the closest following date without missing data. We obtained daily data for each US county from the CDC for the same time period (https://data.cdc.gov/Vaccinations/COVID-19-Vaccinations-in-the-United-States-County/8xkx-amqh). As with the state-level data, we imputed the value for days with missing data by assuming a linear trend. The CDC does not report daily data at the county level for Hawaii or Texas, thus these states are excluded from county-level, time-series analyses.

We aggregated daily estimates of cumulative vaccination to a weekly time scale to match the weekly frequency on which the VSI index is available. We estimated the number of individuals receiving a first dose vaccination in a given week by taking a moving difference of the weekly cumulative vaccination estimates in each county. We excluded from analyses the first week of data in any given county and replaced any negative differences with zeros.

Data on the number of vaccinations by ZIP code are not publicly available from the CDC website, but are reported directly by some US states. Accordingly, we conducted additional analyses at the ZIP code level in two example states that make such data publicly available: California and Texas. The California Department of Public Health provides vaccination rates by ZIP code tabulation area (ZCTA, an approximation of ZIP codes provided by the US Census Bureau) (https://data.ca.gov/dataset/covid-19-vaccine-progress-dashboard-data-by-zip-code), while the Texas Department of State Health Services provides vaccination rates by ZIP code (https://dshs.texas.gov/coronavirus/TexasCOVID19VaccinesbyZIP.xlsx). We converted the Texas ZIP code data to ZCTAs using the latest zip-to-ZCTA crosswalk available from the Uniform Data Systems (UDS) Mapper (https://udsmapper.org/wp-content/uploads/2020/09/Zip_to_zcta_crosswalk_2020.xlsx), in order to more easily link vaccination data with sociodemographic information, which is available for ZCTAs rather than ZIP codes. We additionally obtained data on the cumulative number of vaccinations given by county as of August 11, 2021 from Texas's Department of State Health Services (https://dshs.texas.gov/coronavirus/AdditionalData.aspx).

*Population size and demographics*

We obtained estimates of the age distribution of the population in each county and state from the most recently available 2015-2019 American Community Survey (ACS) (https://data.census.gov/cedsci/table?q=United%20States%20age%20sex&tid=ACSST5Y2019.S0101). The ACS provides estimates of population size within specific age ranges (e.g., ages 5-9 years, 10-14 years, etc). We estimated the number of individuals aged 12 years or over (and thus eligible for vaccination) as the total population of a given region minus the number of people under 10, and minus 40% of the population aged 10-14 years.

We additionally obtained 5-year 2019 ZCTA-level ACS estimates of the age distribution from the IPUMS NHGIS database.[12] The variables selected included information on population proportions by age (percent 65 years or older and percent under 18 years of age), race (percent American Indian or Alaska Native, percent Asian, percent Black or African American, percent Native Hawaiian or Pacific Islander, and percent White), ethnicity (percent Hispanic or Latino [hereafter percent Latinx]), socioeconomic status (percent with college degree or higher, median household income), and health and healthcare (percent without health insurance and percent of persons with a disability).

*Statistical analyses*

In a first analysis we assessed the degree to which search interest in a given week relates to the number of first doses per 100,000 population aged 12 years or older (individuals eligible for vaccination) in the following 1 to 3 weeks. Specifically, for each week between January 11, 2021 through August 22, 2021, we fit a linear mixed-effects regression model to estimate the number of first-dose vaccinations administered over the following 3 weeks per 100,000 eligible individuals in a given county associated with a 10-unit increase in the VSI index. We included a random intercept for state to account for the correlation induced by the nesting of counties within states. In each week, analyses are restricted to the counties that have data on both VSI index and vaccinations, which range from 1,226 to 2,686 (**Supplemental Table 1**).

We next assessed whether search interest early in the course of vaccination rollout in the US is predictive of the cumulative proportion of the population that is vaccinated at a later date. Specifically, we estimated the Pearson correlation coefficient between the mean VSI index in each state or county each month and the percent of the eligible population having received at least 1 vaccine dose in future months. We calculated the monthly mean VSI index based on the month of the first day of each week of data. For example, the VSI index reported for the week of August 30, 2021 contributes to the monthly average for August rather than September.

To visualize how vaccine search interest spatially intersects with vaccination rates, we created bivariate choropleth maps that display the spatial overlap between the VSI index and vaccination rates (as the percent of population aged 12 years or older to have received at least one vaccine dose). This mapping technique allows for an easily interpretable visualization of locations that are relatively high in one variable and low in another, as well as locations that are high or low with respect to both measures. As illustrative examples, we performed separate analyses for California and Texas (using county-level data) and for the Los Angeles and Dallas/Fort Worth metropolitan areas (using ZCTA-level data). We chose these locations based on the availability of ZCTA-level vaccination data and the heterogeneity in vaccination rates. In each analysis, we created a 2x2 grid, resulting in identification of county and ZCTA overlaps by top and bottom tertiles in the following configurations: "high-high" (locations in the top tertile [67th to 100th percentile] of state- or city-specific vaccination rates and the top tertile for search interest), "low-low" (locations in the *bottom* tertiles [0th to 33rd percentile] of both variables), and "high-low" or "low-high" (locations in the top tertile of one variable but bottom tertile of the other). Locations in which one or both variables were either missing or in the middle tertile were classified as "not in high/low or N/A." For these analyses we used the VSI index for the week beginning August 9, 2021. Texas maps were made using vaccination data that were current as of August 11, 2021, and California maps were made using vaccination data that were current as of August 16, 2021. Finally, we report the sociodemographic characteristics of the different groupings of counties and ZCTAs at the intersection of tertiles of the VSI index and vaccination rates, based on the ACS data as described above.

Statistical analyses were conducted in R (version 4.1.1)[13] and maps were created using ArcGIS Pro 2.8.3 (© Esri, Redlands, CA).

# Results

Google search interest for COVID-19 vaccines has varied considerably across location (**Figure 1**) and time (**Figure 2**). In late February 2021, search interest was highest in the Northeastern US, Florida, and the West Coast (**Figure 1A**). In contrast, in mid-July, search interest was highest in Michigan, Missouri, Arkansas, and Louisiana (**Figure 1B**). The time trend of each US state shows that interest was generally high through mid-April and then declined and remained low until late July (**Figure 2A**). Concurrently, there was a rapid increase in the number of first-dose vaccinations administered through approximately the end of April, followed by a sustained but slower rate of first-dose vaccinations through the end of September.

**Figure 1**: County-level Google search interest in vaccines as measured by the Google VSI index for the weeks beginning February 22, 2021 (Panel A) and July 12, 2021 (Panel B).

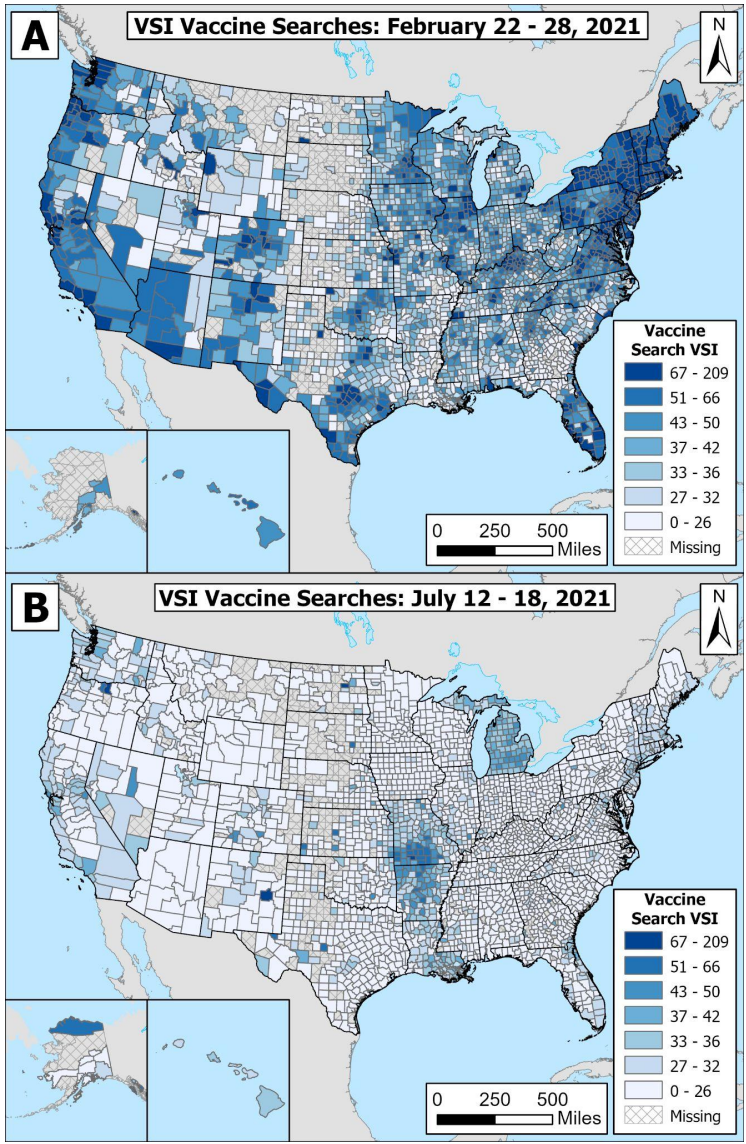

**Figure 2**: State-level Google search interest in vaccines as measured by the Google VSI index (Panel A) and the percentage of the eligible population having received a first dose (Panel B). Each thin line reflects an individual state (*n* = 50); the national level is shown as the thick line in each plot.

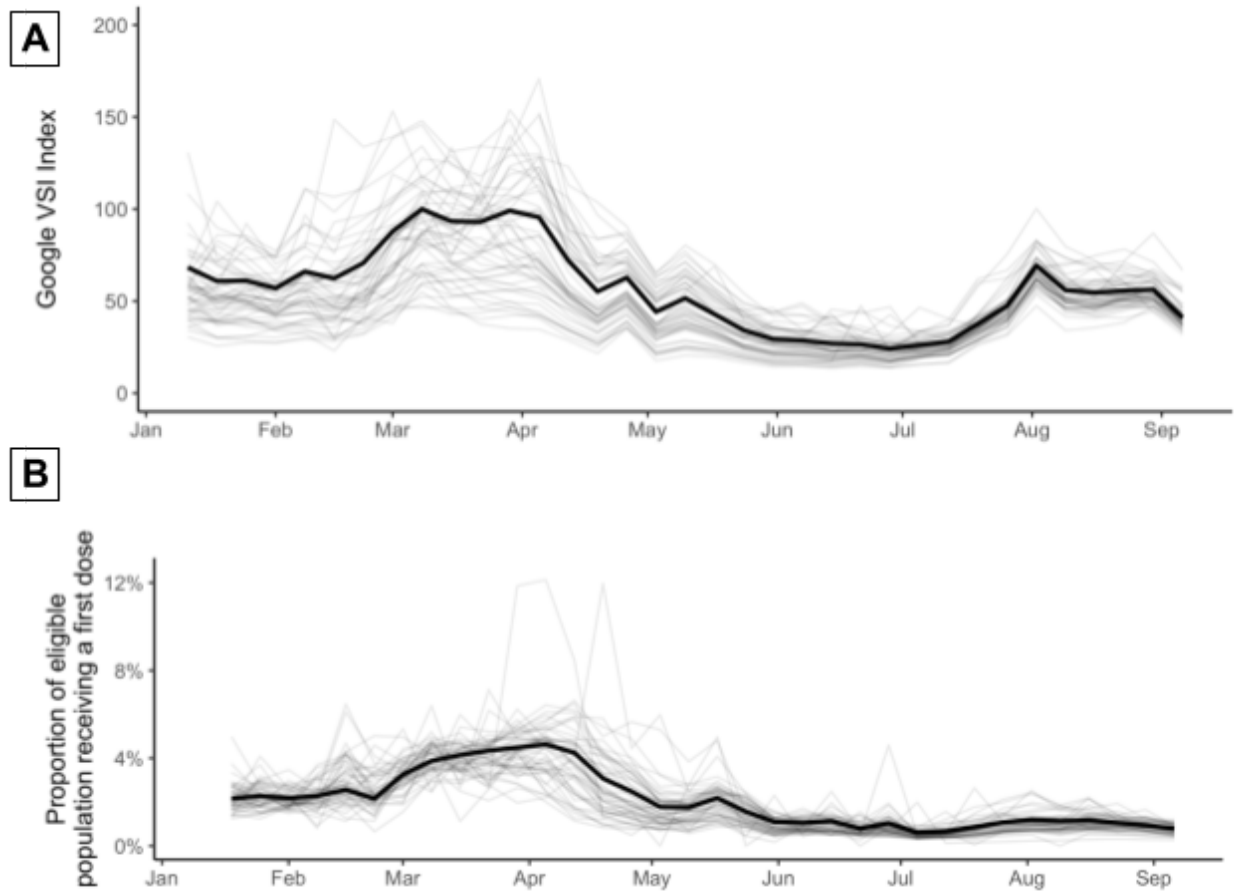

In almost all weeks, the VSI index was positively and statistically significantly associated with the number of first-dose vaccinations per 100,000 eligible individuals in the subsequent 3 weeks (**Figure 3**, see **Supplemental Table 1** for full statistical details). For example, a 10-unit increase in the VSI index based on searches over the week starting on January 11, 2021 was associated with an average of 275.8 (95% confidence interval [CI]: 204.3, 347.4) more first-dose vaccinations per 100,000 eligible individuals in the subsequent 1 to 3 weeks. The strength of the association varied substantially over time, reaching a peak of 947.2 (95% CI: 907.1, 987.3) more first-dose vaccinations over the next 3 weeks per 100,000 eligible individuals the week of March 22, and steadily decreasing to no discernible association the week of August 16, 2021.

The magnitude of the association also varied by state, with the strongest associations between the VSI index and first-dose vaccinations observed in Vermont, Utah, Illinois, and Pennsylvania in April but observed in Colorado, Rhode Island, New York, and Connecticut in July (**Supplemental Figure 1**).

**Figure 3**: Association between county-level values of Google's VSI index and the number of first-dose vaccinations per 100,000 eligible individuals administered in the subsequent 1 to 3 weeks in each county across the United States. A separate mixed-effects model was fit to each week between January 11 and August 22, 2021, inclusive. Estimates from these models reflect the additional number (and 95% confidence intervals) of first-dose vaccinations given in the subsequent 3 weeks associated with a 10-unit change in the county-level VSI index. During this period, the county-level VSI index ranged from 4.40 to 325.78 ($M$ = 40.58, $SD$ = 23.21).

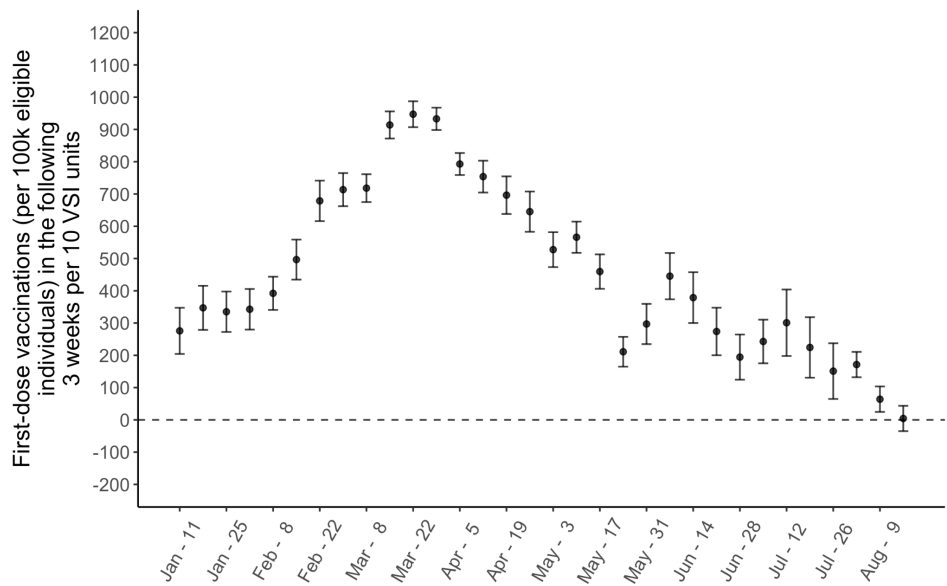

Moreover, the state-level VSI index in the early months (January, February, and March 2021) was positively and statistically significantly correlated with the cumulative vaccination proportion in later months (April - August; **Table 1**, see **Supplemental Table 2** for full statistical details). For example, the correlation between the VSI index in February 2021 and the proportion of the eligible population who have received at least one vaccination dose 6 months later was 0.71 (**Figure 4**). The strongest correlation was between the VSI index in April and cumulative vaccination proportion 1-2 months later ($r \cong 0.89$). Results were similar, though correlations were weaker, when we considered metrics at the county rather than state level (**Supplemental Table 3** and **Supplemental Figure 2**).

**Figure 4**: State-level (*N* = 50) average levels of Google's VSI index in February 2021 are associated with the proportion of eligible individuals in each state having received at least 1 dose of a COVID-19 vaccine at the end of August 2021.

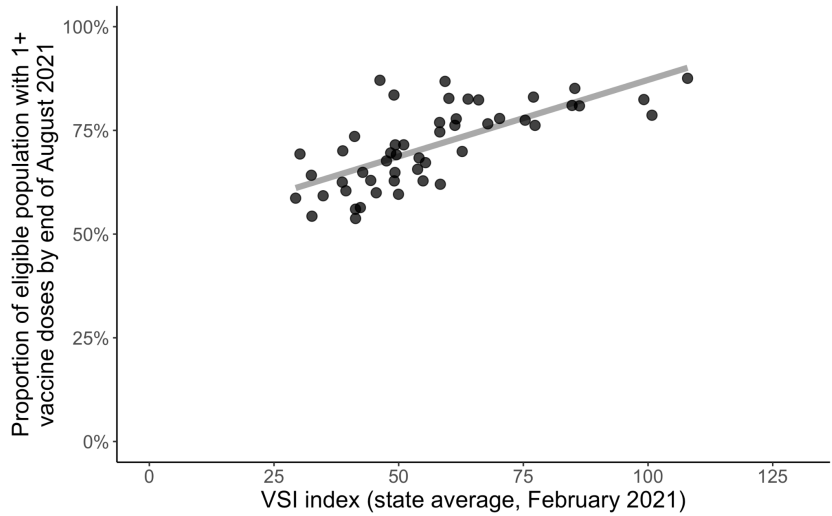

**Table 1**: Pearson correlation coefficient between the monthly average VSI index within each US state (*N* = 50) and percentage of eligible individuals in that state having received at least one dose of a COVID-19 vaccine at varying monthly lags.

|  | Vaccination proportion (percentage of the eligible population with 1+ dose) | | | | | | |
| --- | --- | --- | --- | --- | --- | --- | --- |
| **Mean VSI index in...** | **1 month in the future** | **2 months in the future** | **3 months in the future** | **4 months in the future** | **5 months in the future** | **6 months in the future** | **7 months in the future** |
| **January, 2021** | -0.17 | 0.28* | 0.52*** | 0.56*** | 0.58*** | 0.60*** | 0.62*** |
| **February** | 0.39** | 0.64*** | 0.67*** | 0.69*** | 0.70*** | 0.71*** | |
| **March** | 0.77*** | 0.80*** | 0.80*** | 0.80*** | 0.79*** | | |
| **April** | 0.89*** | 0.89*** | 0.88*** | 0.85*** | | | |
| **May** | 0.86*** | 0.85*** | 0.80*** | | | | |
| **June** | 0.72*** | 0.69*** | | | | | |
| **July** | 0.11 | | | | | | |

\*\*\* $p < .001$, \*\* $p < .01$, \* $p < .05$

Based on the findings above indicating a relationship between the VSI index and subsequent vaccinations, we hypothesized that the VSI index may serve as a leading indicator of vaccine interest and provide novel insights into where new vaccination efforts may be particularly impactful. For example, communities with relatively low vaccination rates but relatively high levels of the VSI index may represent populations where specifically improving vaccine access would be particularly effective for increasing vaccination rates. Even in areas where vaccine availability is very high, the VSI index may help identify locations where people are interested in being vaccinated but where other barriers to access — such as lack of transportation to vaccination sites, or insufficient time off work to either get vaccinated or to recover from potential side effects — could represent a significant deterrent to vaccination. In contrast, communities with low vaccination rates and low VSI index may reflect populations where engagement with information and interest in vaccination might be the primary barrier rather than access to vaccines.

Taking the state of Texas as an example, we categorized the VSI index (**Figure 5A**) and vaccination proportion (**Figure 5B**) as of August 9, 2021 into tertiles relative to the values observed in Texas. Both the VSI index and vaccination proportion varied substantially across the state. Of the 254 counties in Texas, 22 were simultaneously in the highest tertile of the VSI index and lowest tertile of vaccination proportion (**Figure 5C**), perhaps suggesting that higher rates of vaccination could be achieved in these areas specifically through improved access. An additional 23 counties were in the lowest tertile of both the VSI index and vaccination proportion, potentially indicative of communities where existing vaccination outreach programs have been unsuccessful and novel or alternative approaches should be considered.

We repeated this analysis at a finer spatial scale for the Dallas/Fort Worth metropolitan area using metrics at the ZCTA level rather than county level (**Figure 6**). We identified 15 ZCTAs that were simultaneously in the highest tertile of the VSI index and lowest tertile of vaccination proportion (**Figure 6C**), again perhaps suggesting that higher rates of vaccination could be achieved in these areas specifically through improved access. An additional 31 ZCTAs were in the lowest tertile of both the VSI index and vaccination proportion, again potentially suggesting that improved outreach and engagement around vaccination could be beneficial.

We additionally compared the sociodemographic characteristics of ZCTAs classified by the intersection of the VSI index and vaccination proportion (**Table 2**). ZCTAs in the Dallas/Fort Worth area that had simultaneously low vaccination rates and high values of the VSI index have greater population proportions of older adults (ages 65 years and over), residents who identify as White, and persons with a self-reported disability, with lower population proportions of children (ages 17 years and younger) and residents who identify as Black or African American, compared to other VSI/vaccination groupings. By contrast, ZCTAs with low vaccination rates and *low* values of the VSI index included ZCTAs with lower median household incomes, smaller proportions of college graduates, higher proportions of residents who identify as Black or African American, higher proportions of residents who identify as Latinx, and higher proportions of people without health insurance.

We performed analogous analyses for counties across the state of California (**Supplemental Figure 3**) and ZCTAs across the Los Angeles metropolitan area (**Supplemental Figure 4** and **Supplemental Table 4**). As in Texas, the analysis in California highlights counties and ZCTAs where vaccination rates are low but search interest is high, suggesting that higher rates of vaccination may be achieved in these areas through improved access. Additionally, as in Texas, the analysis in California highlights counties and ZCTAs where vaccination rates and search interest are both low, perhaps indicating a need for novel approaches to engagement.

**Figure 5:** Maps of the state of Texas where counties are classified into tertiles of frequency of internet searches related to COVID-19 (as summarized by Google's Vaccine Search Insights [VSI] index for the week of August 9, 2021) within each county (Panel A), tertiles of cumulative proportion of eligible population receiving at least one vaccination as of August 11, 2021 (Panel B), and the cross-classification of these two variables (Panel C).

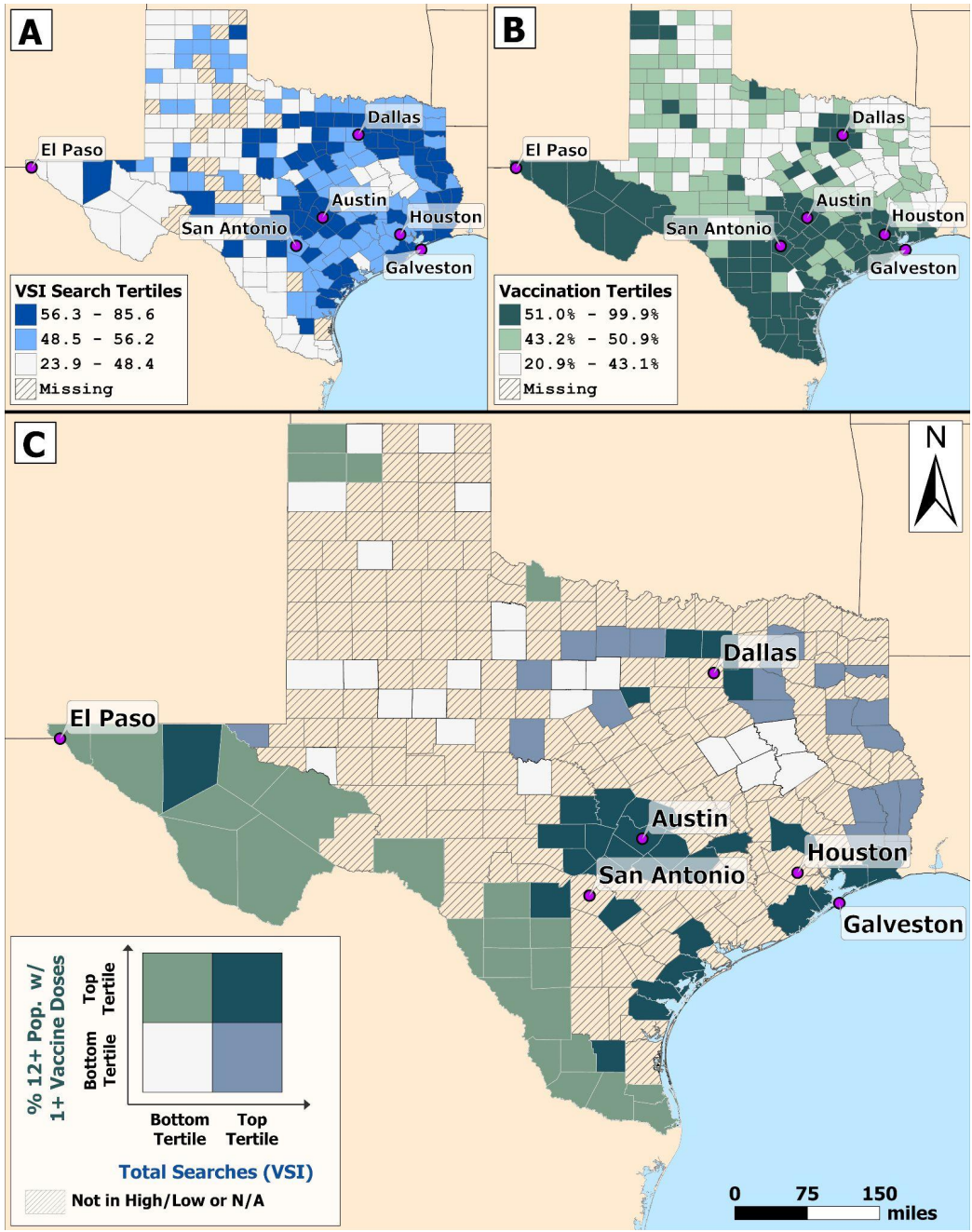

**Figure 6:** Maps of the Dallas/Fort Worth metropolitan area of Texas where ZIP code tabulation areas (ZCTAs) are classified into tertiles of frequency of internet searches related to COVID-19 (as summarized by Google's Vaccine Search Insights [VSI] index for the week of August 9, 2021) within each county (Panel A), tertiles of cumulative proportion of eligible population receiving at least one vaccination as of August 11, 2021 (Panel B), and the cross-classification of these two variables (Panel C).

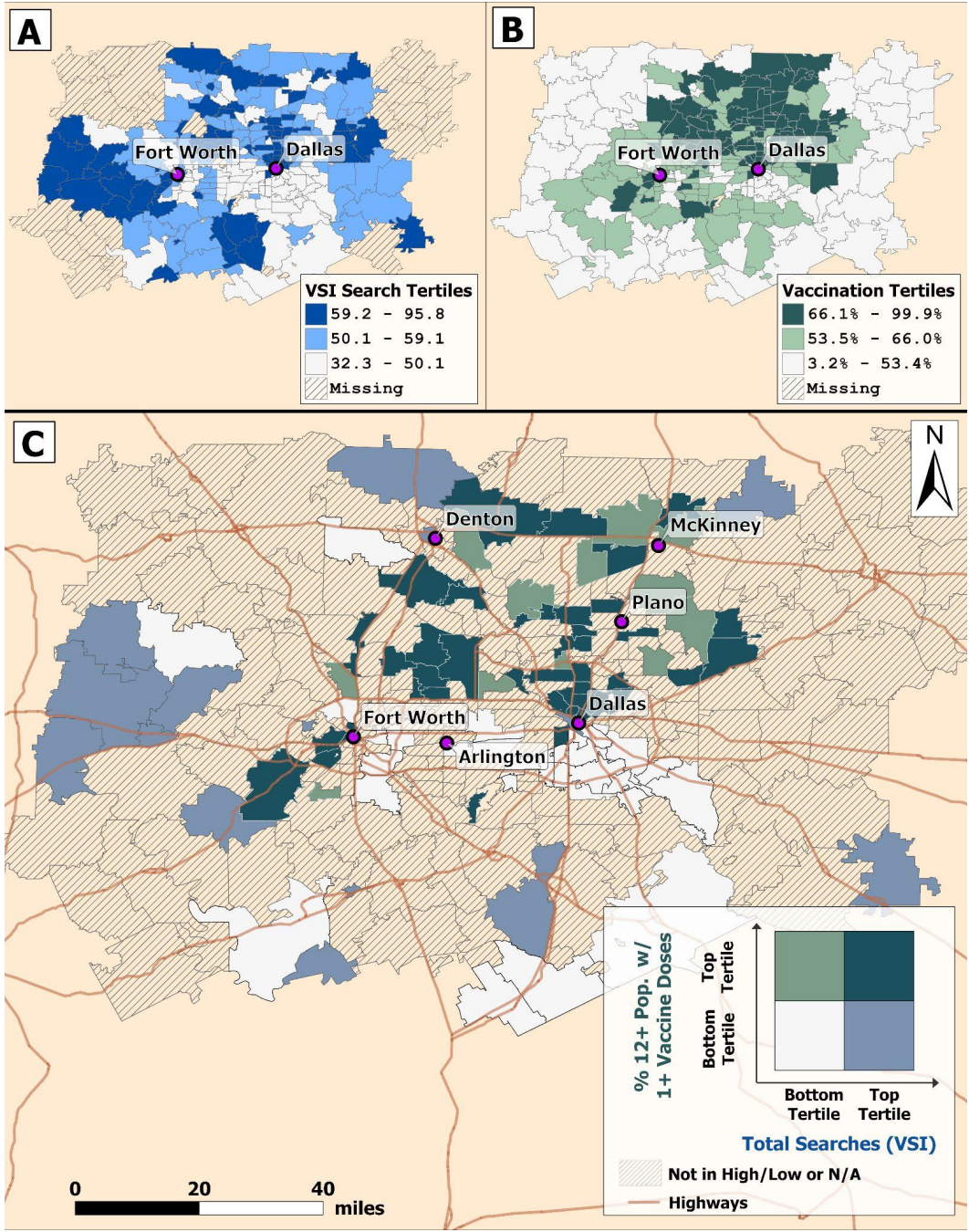

**Table 2**: Sociodemographic characteristics of ZIP code tabulation areas (ZCTAs) in the Dallas/Ft. Worth metropolitan area cross-classified by frequency of internet searches related to COVID-19 (as summarized by Google's Vaccine Search Insights [VSI] index for the week of August 9, 2021) and cumulative proportion of eligible population receiving at least one vaccination as of August 11, 2021. Numbers in cells represent means and standard deviations as reported by the American Community Survey 5-year estimates. "Low" and "High" designations reflect inclusion in bottom and top tertiles of the city-wide distribution, respectively.

| Vaccination Rate | Low | | High | | All Other ZCTAs (n = 172) |
|---|---|---|---|---|---|
| VSI index | High (n = 15) | Low (n = 31) | High (n = 43) | Low (n = 11) | |
| **Age, mean (SD)** | | | | | |
| % Age >64 | 13.75 (6.45) | 10.53 (3.59) | 12.55 (4.56) | 9.06 (2.55) | 11.99 (4.73) |
| % Age <18 | 18.69 (8.71) | 28.41 (3.57) | 21.97 (6.79) | 27.38 (4.13) | 25.04 (5.82) |
| **Race, mean (SD)** | | | | | |
| % American Indian or Alaska Native | 0.89 (1.01) | 0.44 (0.42) | 0.45 (0.38) | 0.54 (0.42) | 0.65 (0.97) |
| % Asian | 1.43 (1.98) | 1.20 (1.36) | 7.79 (7.62) | 12.41 (9.00) | 4.91 (7.29) |
| % Black or African American | 5.73 (9.82) | 31.94 (22.67) | 9.30 (6.47) | 14.17 (10.35) | 12.66 (12.60) |
| % Native Hawaiian or Pacific Islander | 0.66 (2.18) | 0.05 (0.13) | 0.08 (0.29) | 0.18 (0.36) | 0.11 (0.28) |
| % White | 86.14 (12.97) | 57.37 (22.24) | 76.43 (12.52) | 64.83 (14.46) | 73.38 (16.98) |
| **Ethnicity, mean (SD)** | | | | | |
| % Latinx | 16.79 (8.29) | 41.7 (18.4) | 16.84 (11.24) | 22.56 (10.04) | 27.49 (17.87) |
| **Socioeconomic Status, mean (SD)** | | | | | |
| % College Educated | 20.23 (9.23) | 13.19 (5.88) | 55.28 (13.7) | 43.43 (13.91) | 28.24 (15.16) |
| Median Income (USD) | 68,294 (22,552) | 46,176 (12,498) | 101,188 (34,884) | 90,877 (19,826) | 69,353 (24,422) |
| **Health & Healthcare, mean (SD)** | | | | | |
| % Uninsured | 15.82 (6.28) | 24.46 (5.48) | 10.19 (5.40) | 12.95 (5.14) | 17.12 (8.27) |
| % with a Disability | 11.33 (5.35) | 12.72 (3.35) | 7.98 (2.29) | 7.31 (1.58) | 11.16 (5.06) |

# Discussion

Despite ample supply of COVID-19 vaccines in the US, the proportion of the population that has been fully vaccinated remains insufficient across much of the country.[14,15] We evaluated whether Google's VSI index might serve as a useful tool to identify communities with relatively higher interest in vaccination where focused public health efforts might be most successful. We found that between January and August of 2021: (1) Google's weekly VSI index was associated with the number of new vaccinations administered in the subsequent three weeks, and (2) average values of the VSI index in earlier months was strongly correlated with vaccination rates many months later. These results suggest that the VSI index may be useful as a leading indicator of population-level interest in or intent to obtain a COVID-19 vaccine, especially early in the vaccine deployment efforts.

The VSI index was most strongly associated with new vaccinations administered in April versus August 2021, consistent with expectations. In April, the COVID vaccines were still relatively new and vaccination efforts were still ramping up such that vaccine demand exceeded vaccine supply in much of the country. Accordingly, we expect that many people were searching for vaccine information and vaccination appointments. By August 2021, more than 60% of the US population had already received at least one vaccine dose (https://covid.cdc.gov/covid-data-tracker) and many of the remaining individuals were either not eligible (e.g., children under 12 years of age) or had already made up their minds that they did not want to be vaccinated or would only do so if required (https://www.kff.org/coronavirus-covid-19/dashboard/kff-covid-19-vaccine-monitor-dashboard). Given results suggesting that the VSI index may serve as a leading indicator of vaccination interest, we illustrate a graphical approach by which data on the VSI index may be combined with other available data to inform local public health outreach and vaccination efforts.

This work builds on a large and growing body of research that finds that search patterns, based on aggregated and anonymized data, can serve as leading indicators of population-level health outcomes including COVID hospitalizations and death, influenza, Lyme disease, and foodborne illness (https://cloud.google.com/blog/products/ai-machine-learning/google-cloud-is-releasing-the-covid-19-public-forecasts).[4,5,7] However, to the best of our knowledge the current research is the first to show that search patterns may serve as a leading indicator of vaccinations and perhaps a marker of population-level interest in vaccinations in a given place and time. Specifically, many individuals interested in learning more about COVID-19 vaccines or seeking to be vaccinated will search online for this information, such that normalized counts of relevant searches can plausibly serve as an indicator of interest in this topic averaged across a population. The finding that the VSI index is indeed associated with vaccinations in the weeks and months ahead highlights and reflects the importance of online sites and tools for dissemination of information about vaccine safety, efficacy, and availability.

Although availability was a primary barrier to vaccine access in the spring of 2021, that is not likely the predominant barrier to COVID-19 vaccination today, although some vaccine deserts do remain (https://covid19vaccineallocation.org).[16] However, even in areas where vaccine availability is very high, the VSI index may identify locations where people are interested in being vaccinated but where other barriers to access — such as lack of transportation to vaccination sites, or insufficient time off work to either get vaccinated or to recover from potential side effects — could represent a significant deterrent to vaccination. If so, communities with high values of the VSI index and low vaccination rates relative to the larger area may benefit from targeted interventions to make vaccination more convenient or otherwise remove barriers to access. Moreover, the VSI index could be combined with data on vaccine distribution points (e.g., the Vaccine Equity Planner, https://vaccineplanner.org) to gain additional insights into potential barriers to access.

Beyond availability, access, and convenience, Betsch et al.[17] identify confidence, complacency, calculation, and collective responsibility as other factors that contribute to vaccine hesitancy. Relatively higher values of the VSI index in the context of low vaccination rates may indicate that people are potentially interested in vaccination but lack confidence in the available vaccines, the medical system, or policymakers. In such places, focused outreach and engagement by trusted members of the community may be most effective in persuading additional people to be vaccinated. On the other hand, relatively lower values of the VSI index may suggest places with relatively higher levels of complacency (i.e., perceived risk of infection or disease is low, vaccination seen as unnecessary). Locations with low values of the VSI index and low vaccination rates may reflect areas where existing strategies to improve vaccination have been largely unsuccessful in engaging residents and new, different approaches are needed.

While the number of eligible individuals in the US continues to increase, these findings may be relevant to efforts to provide COVID-19 booster shots to eligible individuals or to vaccinate children ages 5-11 years. Beyond the current pandemic, our findings hold relevance for understanding community engagement with public health interventions in the context of future public health emergencies. Data such as the VSI index could be used to rapidly identify areas with heightened interest and willingness to engage around public health interventions, and target limited resources to other areas. Analyses using the VSI data could help identify community-level factors that are associated with reduced engagement in public health interventions and spur local initiatives to address community concerns. In addressing pandemic threats, speed of response is imperative, and metrics such as the VSI index may support efforts to improve communication and intervene where such activities are most needed, saving lives and improving efficiency of response.

These findings need to be interpreted in the context of several important limitations. First, the VSI index is based on anonymized and aggregated search activity and as such provides a measure of average search activity. As with any average, the VSI index may mask important heterogeneity within any given community. Second, the differential privacy algorithms applied to the search data[12] to ensure user privacy increase the variance in the data (i.e., reduce the signal to noise ratio), particularly in areas with smaller populations. Thus, we expect the VSI index to be more robust in counties and ZIP codes with relatively large populations. Third, the VSI index only reflects the search activity of Google users searching in English and may not represent the interests or attitudes of individuals without reliable internet access, those searching in other languages, or those who use other search providers. Fourth, the association between the VSI index and new vaccinations varies across time and place. This suggests that any actions based on the VSI index must be based on recent data and recognize the potential of changing circumstances over time. Finally, we note that the unit of analysis here is at the level of the county or ZIP code. Inferences and associations observed at the aggregate level may not apply to individuals. Likewise, while many individuals searching for vaccine information may themselves be unvaccinated and seeking vaccination, search behaviors are diverse and complex. For example, individuals may search for information for others rather than themselves or may perform searches in a location different from where they live. Moreover, search behavior is expected to vary over time in response to, for example, media coverage, interest in or concerns about booster shots, or shifting eligibility criteria.

      Notwithstanding the above limitations, these findings represent a novel use of anonymized and aggregated internet search data as a leading indicator of COVID-19 vaccination. Our analyses show that the VSI index, in combination with other available data, may provide actionable insights to public health officials and community leaders seeking to increase local rates of vaccinations. Whether these insights contribute meaningfully to the public health response to the current COVID-19 pandemic remains an important unresolved question, which will in part depend on the future course of the pandemic. Beyond the near-term public health benefits, insights gained from our analysis of the VSI index may also serve as valuable,

early indicators of the population's information needs or willingness to engage around public health interventions in the context of future emergencies, improving effectiveness and efficiency of local response efforts.

## Data Availability

All data used in this paper are publicly available, as described in the Methods section.

## Acknowledgements

We are grateful to Elżbieta Brzóz, Bruno Delmonte, Tetiana Kedzierska, Jan Machowski, Don Metzler, Sarah Montgomery-Taylor, Arti Patankar, and Chris Scott, for their help and advice.

## Competing Interests

Dr. Wellenius serves as a consultant to Google, LLC (Mountain View, CA). This work was funded in part by an unrestricted gift from Google, LLC to Boston University School of Public Health. In addition, Google, LLC provided support in the form of salaries for employees. Several of the study authors were also involved in the development and launch of Google's Vaccination Search Insights. Google did not have any additional role in the study design, data collection and analysis, or preparation of the manuscript. All manuscripts co-authored by employees of Google are reviewed prior to journal submission to ensure that they meet Google's standards.

# Author Contributions

SM, KRS, JHL, KJL, SB, YS, AB, CS, KS, TS, KC, JIL, AAS, EG, GAW developed the concept. MS, SB, CK, AK, JG, TG, AB, MY, YM computed the data. SM, MS, KRS, JHL, KJL, JIL, AAS, EG, GAW designed the study. SM, MS, KRS, KJL, YS analyzed the data and performed statistical analysis. All authors contributed to writing the manuscript.

# Supplementary Information

**Supplemental Table 1**: Estimates from linear mixed-effects regression models estimating the number of first-dose vaccinations administered over the following 3 weeks per 100,000 eligible individuals in a given county associated with a 10-unit increase in the VSI index. Models included a random intercept for state to account for the correlation induced by the nesting of counties within states

| Week | Term | b | SE | CI lower | CI upper | n counties | n states |
|---:|---|---:|---:|---:|---:|---:|---:|
| 1/11/21 | Intercept | 4816.22 | 350.54 | 4129.15 | 5503.28 | 1226 | 48 |
|  | VSI index | 275.83 | 36.51 | 204.28 | 347.39 |  |  |
| 1/18/21 | Intercept | 4874.14 | 352.36 | 4183.52 | 5564.76 | 1765 | 48 |
|  | VSI index | 347.14 | 34.84 | 278.85 | 415.43 |  |  |
| 1/25/21 | Intercept | 4421.52 | 348.52 | 3738.42 | 5104.62 | 2306 | 48 |
|  | VSI index | 335.09 | 31.95 | 272.46 | 397.72 |  |  |
| 2/1/21 | Intercept | 4603.15 | 347.21 | 3922.63 | 5283.68 | 2541 | 48 |
|  | VSI index | 342.73 | 32.1 | 279.82 | 405.64 |  |  |
| 2/8/21 | Intercept | 4676.61 | 349.54 | 3991.51 | 5361.71 | 2602 | 48 |
|  | VSI index | 392.17 | 26.25 | 340.73 | 443.61 |  |  |
| 2/15/21 | Intercept | 6068.39 | 438.2 | 5209.52 | 6927.26 | 2606 | 48 |
|  | VSI index | 496.67 | 31.64 | 434.67 | 558.68 |  |  |
| 2/22/21 | Intercept | 5874.2 | 412.76 | 5065.18 | 6683.21 | 2627 | 48 |
|  | VSI index | 678.64 | 32 | 615.91 | 741.37 |  |  |
| 3/1/21 | Intercept | 5362.13 | 405.26 | 4567.83 | 6156.44 | 2669 | 48 |
|  | VSI index | 713.47 | 26.19 | 662.14 | 764.8 |  |  |
| 3/8/21 | Intercept | 4106.12 | 326.66 | 3465.87 | 4746.37 | 2686 | 48 |
|  | VSI index | 718.08 | 22 | 674.95 | 761.21 |  |  |
| 3/15/21 | Intercept | 3690.9 | 338.5 | 3027.44 | 4354.37 | 2665 | 48 |
|  | VSI index | 913.9 | 21.46 | 871.84 | 955.96 |  |  |
| 3/22/21 | Intercept | 3120.84 | 298.85 | 2535.09 | 3706.58 | 2667 | 48 |
|  | VSI index | 947.18 | 20.46 | 907.08 | 987.28 |  |  |
| 3/29/21 | Intercept | 1945.96 | 294.05 | 1369.62 | 2522.29 | 2655 | 48 |
|  | VSI index | 932.87 | 17.62 | 898.34 | 967.41 |  |  |
| 4/5/21 | Intercept | 822.8 | 239.24 | 353.9 | 1291.7 | 2647 | 48 |
|  | VSI index | 792.86 | 17.39 | 758.78 | 826.94 |  |  |

**Supplemental Table 1** *Continued*

| Week | Term | b | SE | CI lower | CI upper | n counties | n states |
|---:|---|---:|---:|---:|---:|---:|---:|
| 4/12/21 | Intercept | 1407.2 | 549.21 | 330.74 | 2483.66 | 2620 | 48 |
| | VSI index | 753.8 | 25.15 | 704.5 | 803.1 | | |
| 4/19/21 | Intercept | 1695.45 | 596.38 | 526.55 | 2864.34 | 2561 | 48 |
| | VSI index | 696.35 | 29.71 | 638.11 | 754.59 | | |
| 4/26/21 | Intercept | 1325.56 | 675.42 | 1.73 | 2649.38 | 2634 | 48 |
| | VSI index | 645.22 | 31.83 | 582.82 | 707.62 | | |
| 5/3/21 | Intercept | 2111.91 | 211.79 | 1696.79 | 2527.02 | 2527 | 48 |
| | VSI index | 527.55 | 27.62 | 473.41 | 581.69 | | |
| 5/10/21 | Intercept | 1179.71 | 184.02 | 819.03 | 1540.39 | 2559 | 48 |
| | VSI index | 566 | 24.7 | 517.6 | 614.4 | | |
| 5/17/21 | Intercept | 983.78 | 168.08 | 654.34 | 1313.22 | 2549 | 48 |
| | VSI index | 459.53 | 27.22 | 406.18 | 512.89 | | |
| 5/24/21 | Intercept | 1558.69 | 116.65 | 1330.06 | 1787.32 | 2508 | 48 |
| | VSI index | 211.1 | 23.58 | 164.88 | 257.32 | | |
| 5/31/21 | Intercept | 1455.96 | 132.45 | 1196.36 | 1715.57 | 2455 | 48 |
| | VSI index | 297.09 | 31.83 | 234.71 | 359.47 | | |
| 6/7/21 | Intercept | 1239.4 | 152.12 | 941.25 | 1537.54 | 2449 | 48 |
| | VSI index | 445.35 | 36.62 | 373.58 | 517.12 | | |
| 6/14/21 | Intercept | 1243.84 | 157.27 | 935.58 | 1552.1 | 2425 | 48 |
| | VSI index | 378.84 | 40.2 | 300.05 | 457.62 | | |
| 6/21/21 | Intercept | 1282.66 | 156.39 | 976.13 | 1589.19 | 2461 | 48 |
| | VSI index | 273.91 | 37.58 | 200.24 | 347.57 | | |
| 6/28/21 | Intercept | 1338.2 | 149.1 | 1045.96 | 1630.43 | 2421 | 48 |
| | VSI index | 194.46 | 35.7 | 124.48 | 264.44 | | |
| 7/5/21 | Intercept | 1488.62 | 155.05 | 1184.72 | 1792.52 | 2496 | 48 |
| | VSI index | 242.92 | 34.5 | 175.3 | 310.54 | | |
| 7/12/21 | Intercept | 1966.65 | 326.71 | 1326.3 | 2607.01 | 2533 | 48 |
| | VSI index | 300.96 | 52.55 | 197.96 | 403.97 | | |
| 7/19/21 | Intercept | 2168.72 | 316.01 | 1549.35 | 2788.09 | 2584 | 48 |
| | VSI index | 224.43 | 47.84 | 130.67 | 318.2 | | |
| 7/26/21 | Intercept | 2471.35 | 326.93 | 1830.56 | 3112.13 | 2634 | 48 |
| | VSI index | 151.13 | 44.03 | 64.83 | 237.42 | | |

**Supplemental Table 1** *Continued*

| Week | Term | b | SE | CI lower | CI upper | n counties | n states |
|---|---|---:|---:|---:|---:|---:|---:|
| 8/2/21 | Intercept | 1728.49 | 196.32 | 1343.7 | 2113.28 | 2654 | 48 |
|  | VSI index | 171.41 | 20.04 | 132.14 | 210.69 |  |  |
| 8/9/21 | Intercept | 2440.82 | 178.59 | 2090.77 | 2790.86 | 2639 | 48 |
|  | VSI index | 64.09 | 20.14 | 24.61 | 103.57 |  |  |
| 8/16/21 | Intercept | 2352.1 | 152.54 | 2053.12 | 2651.08 | 2633 | 48 |
|  | VSI index | 4.3 | 20.04 | -34.98 | 43.57 |  |  |

**Supplemental Figure 1**: To assess whether the relationship between search interest and first dose vaccinations varied by state in April and July, we estimated a series of linear models using a generalized estimating equation in which the number of new vaccinations in the following 3 weeks (per 100,000 eligible individuals) was regressed on VSI index (in 10s of units) and week. A separate model was estimated for each state in April and July. Data were clustered at the county level and an independence correlation structure was specified. Estimates reflect how many first dose vaccinations (per 100,000 eligible individuals) were associated with a 10-unit change in VSI index per week and county. Error bars reflect +/- 2 robust *SE*s. States names are colored by quintile of their April point estimate to facilitate comparing rank order between April and July.

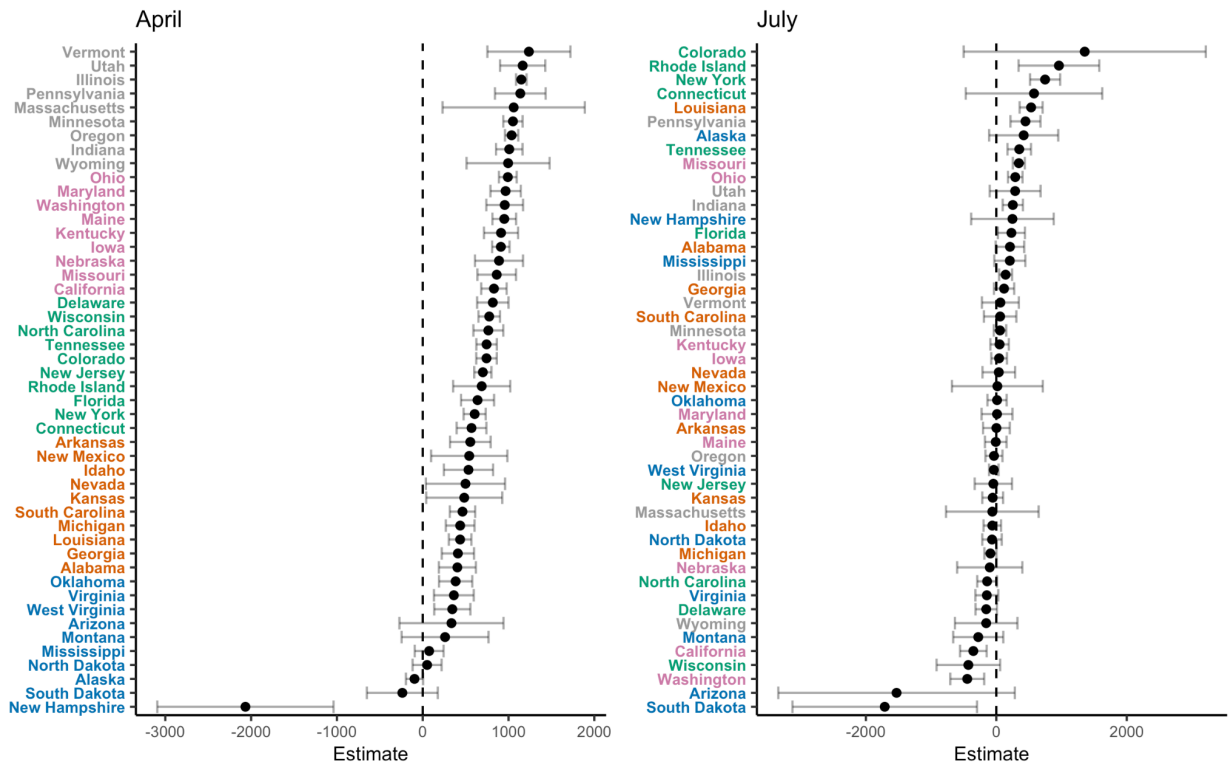

**Supplemental Table 2**: Pearson correlation coefficient between the monthly average VSI index within each US state (*N* = 50) and percentage of the eligible state population having received at least one dose of a COVID-19 vaccine at varying monthly lags.

| VSI index in | Vaccination percentage | *Pearon's r* | *p* | *n* |
|---|---|---|---|---|
| January | 1 month in future | -0.170 | 2.37E-01 | 50 |
| | 2 months in future | 0.280 | 4.90E-02 | 50 |
| | 3 months in future | 0.516 | 1.25E-04 | 50 |
| | 4 months in future | 0.561 | 2.27E-05 | 50 |
| | 5 months in future | 0.583 | 8.98E-06 | 50 |
| | 6 months in future | 0.600 | 4.17E-06 | 50 |
| | 7 months in future | 0.620 | 1.60E-06 | 50 |
| February | 1 month in future | 0.389 | 5.23E-03 | 50 |
| | 2 months in future | 0.640 | 5.59E-07 | 50 |
| | 3 months in future | 0.672 | 9.08E-08 | 50 |
| | 4 months in future | 0.689 | 3.16E-08 | 50 |
| | 5 months in future | 0.698 | 1.71E-08 | 50 |
| | 6 months in future | 0.705 | 1.10E-08 | 50 |
| March | 1 month in future | 0.774 | 4.30E-11 | 50 |
| | 2 months in future | 0.799 | 3.50E-12 | 50 |
| | 3 months in future | 0.804 | 2.14E-12 | 50 |
| | 4 months in future | 0.804 | 2.03E-12 | 50 |
| | 5 months in future | 0.789 | 1.05E-11 | 50 |
| April | 1 month in future | 0.888 | 8.30E-18 | 50 |
| | 2 months in future | 0.888 | 7.76E-18 | 50 |
| | 3 months in future | 0.881 | 3.29E-17 | 50 |
| | 4 months in future | 0.849 | 6.78E-15 | 50 |
| May | 1 month in future | 0.863 | 8.35E-16 | 50 |
| | 2 months in future | 0.848 | 7.65E-15 | 50 |
| | 3 months in future | 0.802 | 2.44E-12 | 50 |
| June | 1 month in future | 0.717 | 4.64E-09 | 50 |
| | 2 months in future | 0.687 | 3.60E-08 | 50 |
| July | 1 month in future | 0.114 | 4.32E-01 | 50 |

**Supplemental Table 3**: Pearson correlation coefficient between the monthly average VSI index at the county level and percentage of the eligible county population having received at least one dose of a COVID-19 vaccine at varying monthly lags.

| VSI index in | Vaccination percentage | Pearon's r | p | n |
|---|---|---|---|---|
| January | 1 month in future | 0.092 | 8.57E-06 | 2313 |
| | 2 months in future | 0.271 | 4.47E-40 | 2313 |
| | 3 months in future | 0.402 | 2.25E-90 | 2312 |
| | 4 months in future | 0.431 | 8.59E-105 | 2300 |
| | 5 months in future | 0.427 | 3.67E-102 | 2287 |
| | 6 months in future | 0.410 | 6.73E-93 | 2273 |
| | 7 months in future | 0.375 | 1.26E-75 | 2235 |
| February | 1 month in future | 0.307 | 5.00E-60 | 2707 |
| | 2 months in future | 0.445 | 6.60E-131 | 2694 |
| | 3 months in future | 0.477 | 3.16E-150 | 2645 |
| | 4 months in future | 0.474 | 1.16E-146 | 2619 |
| | 5 months in future | 0.454 | 5.09E-132 | 2596 |
| | 6 months in future | 0.411 | 3.26E-104 | 2536 |
| March | 1 month in future | 0.462 | 3.90E-145 | 2742 |
| | 2 months in future | 0.513 | 1.29E-180 | 2686 |
| | 3 months in future | 0.516 | 1.18E-180 | 2660 |
| | 4 months in future | 0.491 | 3.72E-160 | 2633 |
| | 5 months in future | 0.452 | 2.29E-129 | 2565 |
| April | 1 months in future | 0.483 | 3.12E-157 | 2686 |
| | 2 months in future | 0.491 | 7.98E-162 | 2663 |
| | 3 months in future | 0.464 | 2.95E-141 | 2642 |
| | 4 months in future | 0.421 | 2.71E-111 | 2574 |
| May | 1 month in future | 0.439 | 1.67E-125 | 2652 |
| | 2 months in future | 0.421 | 7.73E-114 | 2642 |
| | 3 months in future | 0.376 | 1.51E-87 | 2577 |
| June | 1 month in future | 0.254 | 1.36E-39 | 2605 |
| | 2 months in future | 0.242 | 1.03E-35 | 2589 |
| July | 1 month in future | 0.068 | 4.59E-04 | 2657 |

**Supplemental Figure 2:** County-level average VSI index in February and proportion of eligible county population with at least 1 dose of a COVID-19 vaccine at the end of August.

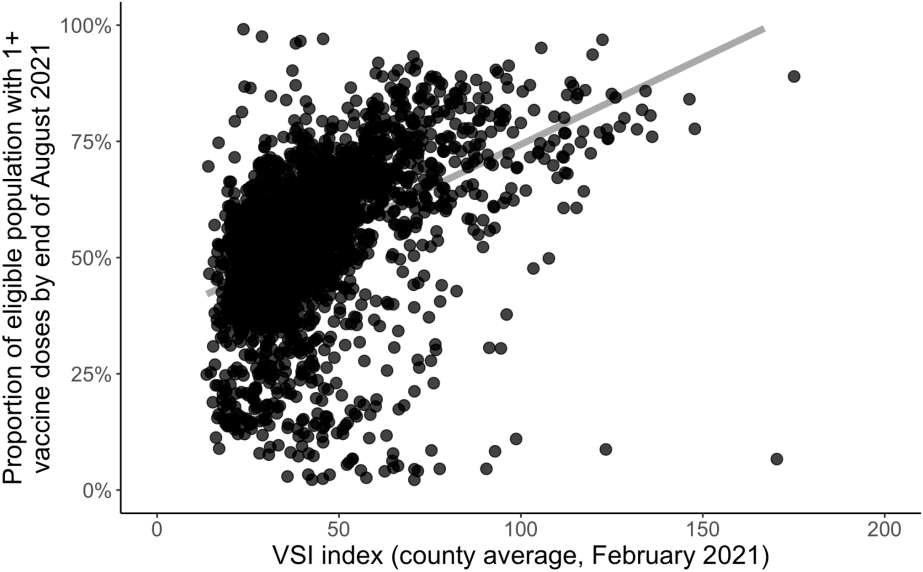

**Supplemental Figure 3:** Maps of the state of California where counties are classified into tertiles of frequency of internet searches related to COVID-19 (as summarized by Google's Vaccine Search Insights [VSI] index for the week of August 9, 2021) within each county (Panel A), tertiles of cumulative proportion of eligible population receiving at least one vaccination as of August 16, 2021 (Panel B), and the cross-classification of these two variables (Panel C).

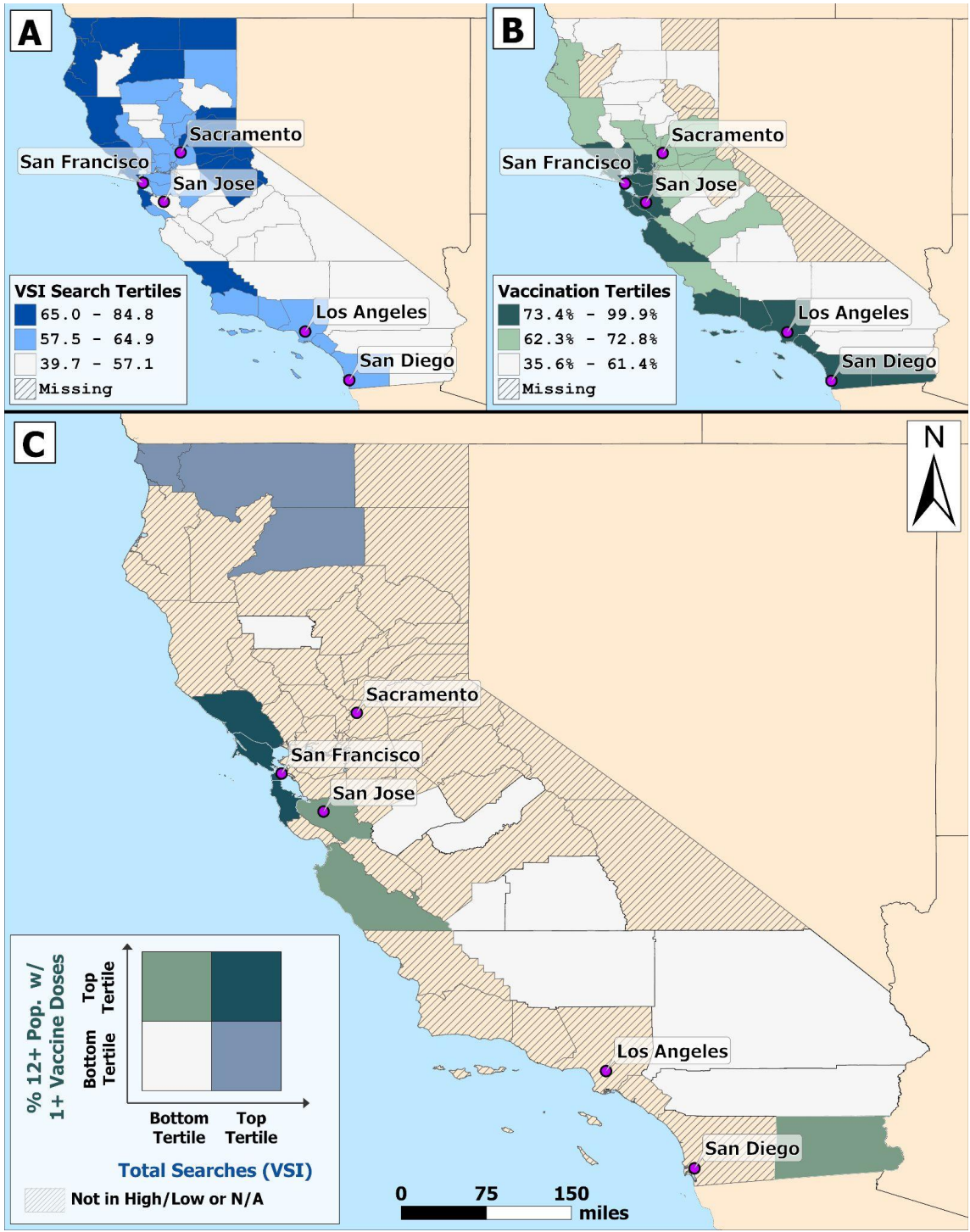

**Supplemental Figure 4.** Maps of the Los Angeles metropolitan area of California where ZIP code tabulation areas (ZCTAs) are classified into tertiles of frequency of internet searches related to COVID-19 (as summarized by Google's Vaccine Search Insights [VSI] index for the week of August 9, 2021) within each county (Panel A), tertiles of cumulative proportion of eligible population receiving at least one vaccination as of August 16, 2021 (Panel B), and the cross-classification of these two variables (Panel C).

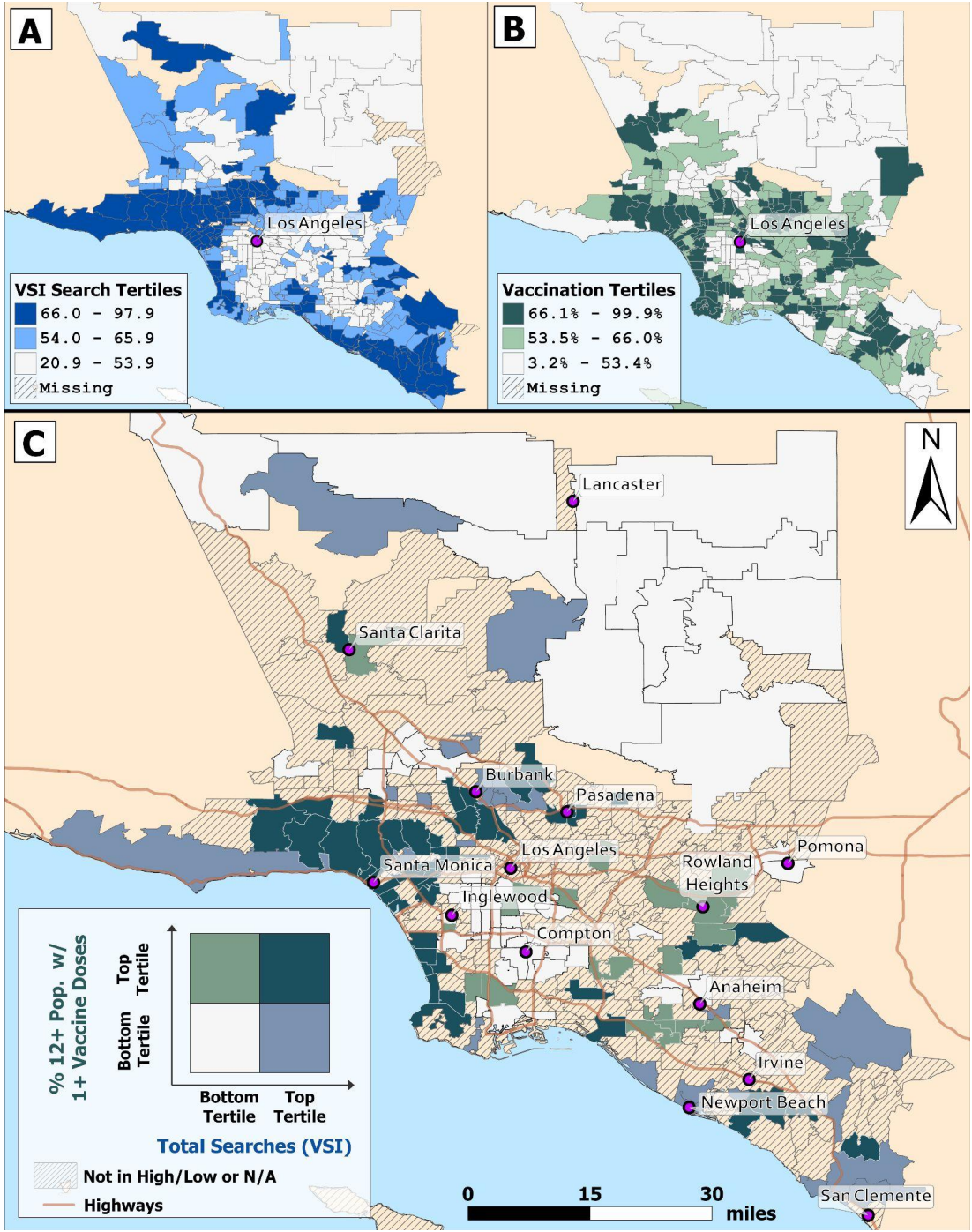

**Supplemental Table 4:** Sociodemographic characteristics of ZIP code tabulation areas (ZCTAs) in the Los Angeles metropolitan area cross-classified by frequency of internet searches related to COVID-19 (as summarized by Google's Vaccine Search Insights [VSI] index for the week of August 9, 2021) and cumulative proportion of eligible population receiving at least one vaccination as of August 16, 2021. Numbers in cells represent means and standard deviations as reported by the American Community Survey 5-year estimates.

| Vaccination Rate | Low | | High | | All Other ZCTAs (n = 229) |
|---|---|---|---|---|---|
| **VSI index** | High (n = 26) | Low (n = 54) | High (n = 51) | Low (n = 20) | |
| **Age** | | | | | |
| % Age >64 | 16.11 (5.38) | 11.22 (4.83) | 18.53 (10.89) | 14.32 (4.97) | 15.35 (8.63) |
| % Age <18 | 17.66 (4.76) | 25.76 (5.03) | 18.56 (6.48) | 23.08 (5.72) | 19.48 (5.99) |
| **Race** | | | | | |
| % American Indian or Alaska Native | 0.35 (0.34) | 1.06 (1.76) | 0.40 (0.45) | 0.63 (0.37) | 0.60 (0.59) |
| % Asian | 10.21 (8.70) | 5.84 (6.02) | 15.64 (10.06) | 33.44 (22.90) | 18.12 (14.67) |
| % Black or African American | 2.29 (1.83) | 15.29 (16.72) | 3.40 (2.68) | 3.40 (2.78) | 5.76 (9.28) |
| % Native Hawaiian or Pacific Islander | 0.19 (0.29) | 0.29 (0.43) | 0.09 (0.11) | 0.45 (0.72) | 0.38 (1.14) |
| % White | 74.83 (11.65) | 47.18 (16.23) | 71.02 (12.77) | 40.14 (18.43) | 56.02 (18.77) |
| **Ethnicity** | | | | | |
| % Latinx | 23.02 (13.25) | 63.56 (19.90) | 14.67 (8.40) | 45.90 (26.14) | 37.12 (23.91) |
| **Socioeconomic Status** | | | | | |
| % College Educated | 45.82 (14.90) | 15.50 (8.10) | 63.10 (10.49) | 27.85 (14.32) | 38.56 (17.17) |
| Median Income (USD) | 88,333.92 (28,954.15) | 54,701.85 (12,727.5) | 115,106.24 (37,935.39) | 71,185.55 (22,865.38) | 81,474.43 (27,857.27) |
| **Health & Healthcare** | | | | | |
| % Uninsured | 7.04 (3.68) | 11.94 (3.76) | 4.17 (2.84) | 8.69 (3.85) | 8.02 (4.95) |
| % with a Disability | 9.75 (3.22) | 10.98 (3.52) | 8.63 (3.59) | 9.42 (1.69) | 10.13 (5.61) |